\documentclass[twocolumn,aps,superscriptaddress,Xlinenumbers]{revtex4-1}
\usepackage[T1]{fontenc}
\usepackage[utf8]{inputenc}
\usepackage{color}
\usepackage{amsmath}
\usepackage{amssymb}
\usepackage{graphicx}
\usepackage[unicode=true,pdfusetitle,
 bookmarks=true,bookmarksnumbered=false,bookmarksopen=false,
 breaklinks=false,pdfborder={0 0 1},backref=false,colorlinks=true]
 {hyperref}
\hypersetup{
 citecolor=blue}

\makeatletter
\graphicspath{{./}{./Figs/}} 

\makeatother

\begin{document}
\title{}
\title{Aberration-driven tilted emission in degenerate cavities }
\author{S. V. Gurevich}
\affiliation{Institute for Theoretical Physics, University of Münster Wilhelm-Klemm-Str. 9,
D-48149 Münster, Germany}
\affiliation{Departament de Física and IAC$^{3}$, Universitat de les Illes Balears,
C/ Valldemossa km 7.5, 07122 Mallorca, Spain}
\author{F. Maucher}
\affiliation{Departament de Física and IAC$^{3}$, Universitat de les Illes Balears,
C/ Valldemossa km 7.5, 07122 Mallorca, Spain}
\affiliation{Faculty of Mechanical, Maritime and Materials Engineering; Department
of Precision and Microsystems Engineering, Delft University of Technology,
2628 CD, Delft, The Netherlands}
\author{J. Javaloyes}
\affiliation{Departament de Física and IAC$^{3}$, Universitat de les Illes Balears,
C/ Valldemossa km 7.5, 07122 Mallorca, Spain}
\begin{abstract}
The compensation of chromatic dispersion opened new avenues and extended
the level of control upon pattern formation in the \textit{temporal
domain}. In this manuscript, we propose the use of a nearly-degenerate
laser cavity as a general framework allowing for the exploration of
higher contributions to diffraction in the \textit{spatial} domain.
Our approach leverages the interplay between optical aberrations and
the proximity to the self-imaging condition which allows to cancel
or reverse paraxial diffraction. As an example, we show how spherical
aberrations materialize into a transverse bilaplacian operator and,
thereby, explain the stabilization of temporal solitons travelling
off-axis in an unstable mode-locked broad-area surface-emitting laser.
We disclose an analogy between these regimes and the dynamics of a
quantum particle in a double well potential.
\end{abstract}
\maketitle
The understanding of self-organized spatiotemporal patterns is key
in photonics and the formation of shocks, vortices, tilted waves,
cross-roll patterns, weak optical turbulence, and localized structures
was observed experimentally and studied theoretically in large-aspect-ratio
lasers, see e.g.,~\citep{L-CSF-94,Lega_1994,HMT-PRL-95,SSW-PRL-97,TSW-PRL-98,ABR-PR-99,BTB-NAT-02,AFO-AAM-09,GBG-PRL-10,J-PRL-16,GPTT-PRL-21}.
Another recent example of multidimensional spatio-temporal self-organization
is the experimental observation of spatiotemporal mode-locking in
multimode optical fibers~\citep{WSP-Nature-20,DXL-PRL-22,WRC-Optica-22}.

Dispersion compensation consists in combining elements with opposed
chromatic properties to achieve an overall partial or total cancellation
of the second order dispersion. This simple yet powerful idea permitted
exploring the influence of higher order contributions. While optical
temporal localized structures often result from the balance between
self-phase modulation and anomalous second-order dispersion~\citep{LCK_NP_10,HBJ-NAP-14,MJB-PRL-14},
it was recently proven that third and fourth-order dispersion lead
to unforeseen effects such as the stabilization of solitons and frequency
combs \citep{TG-OL-10,PGL-OL-14,TABB-PRA-20}, symmetry breaking \citep{LMK-PRL-13},
the control of modulational instabilities \citep{TML-OL-07,DKB-OL-13}
or the realization of purely quartic solitons \citep{BSS-NAC-16,DRH-APLP-21}
as predicted in \citep{KH-OC-94}. 

The paraxial diffraction emerging as a beam propagate is mathematically
equivalent to that of second-order chromatic dispersion. Optical cavities
in which the path of light is folded onto itself may contain a transverse
plane that is its own image after a round-trip \citep{A-APO-69,SIEGMAN-BOOK}.
This so-called stigmatic condition is equivalent to an effective cancellation
of paraxial diffraction. Imposing additionally the nullification of
the round-trip wavefront curvature does not only achieve the self-imaging
condition (SIC) for the field intensity, but also for its \emph{amplitude}.
Self-imaging cavities have attracted a great deal of attentions for
their rich spatio-temporal dynamics \citep{TSW-PRL-98,HBE-OE-08,LKF-OE-10,J-PRL-16,BVM-Optica-22,NathanPHD}
and for their application in speckle-free imaging \citep{KLR-Optica-16,MEY-NNP-21},
frequency comb multiplexing \citep{PWC-Optica-22}, controllable and
reconfigurable multimode fields \citep{EMF-PRL-22,BVM-PR-23}, see
\citep{CCB-NRP-19} for a review. The proximity of the SIC was used
to manipulate the spatial coherence of the field \citep{TVF-JOSAA-91,MEY-NNP-21,EMF-PRL-22,KBJ-APLP-22,POT-NAP-22},
create a perfect coherent absorber \citep{SWH-SCI-22}, realising
topological band structures \citep{YZH-NATCOM-22}, or to form propagation
invariant beams \citep{CBT-OE-18}. Aberrations become crucial as
the SIC is approached~\citep{Clark2020,PhysRevA.104.013524,BVM-Optica-22},
yet, their effect on spatio-temporal laser dynamics received comparatively
less attention. 
\begin{figure}[b]
\includegraphics[viewport=0bp 0bp 347bp 242bp,clip,width=1\columnwidth]{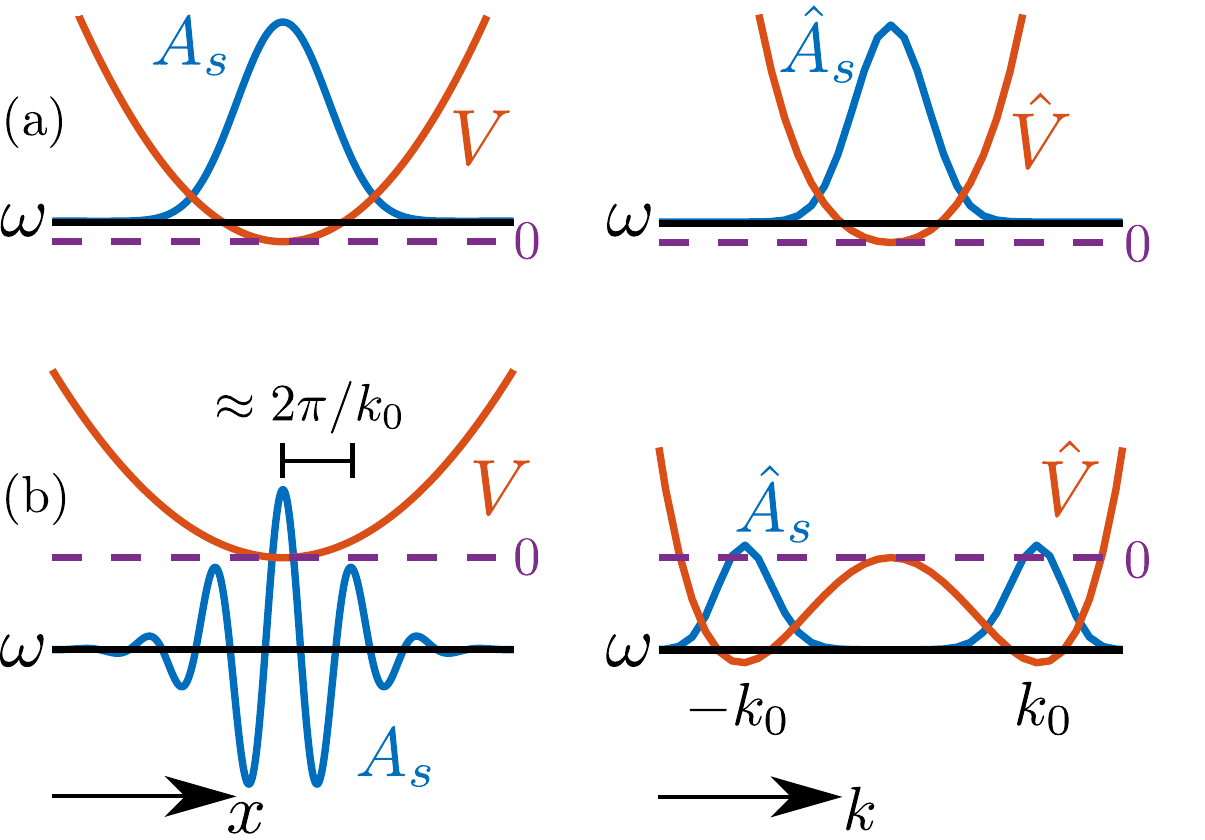}
\caption{Schematic of the fundamental Hermite-Gauss (HG) mode $A_{s}(x)$ (blue)
and the corresponding potential $V(x)$ (orange) of Eq.~\eqref{eq:funny_oscillator}
in the real (left) and Fourier space (right) for (a) $b<0$, $c>0$
and $s=1$ (b) $b>0$, $c>0$ and $s=1$. The resulting double-well
potential in the Fourier space with the minima located at $k_{0}$
shown in (b) correspond to a tilted HG mode in the real space.}
\label{fig1}
\end{figure}
Photonic crystals also allow for dispersion control \citep{VBH-Nature-05,FLFP-OE-06}
and lead to the zero-diffraction regime \citep{SSH-PRE-06}. Higher
order spatial operators occur in fiber lasers \citep{KZC-PhotRes-21}
and in optical cavities either at the onset of bistability \citep{TML-PRL-94}
or containing a photorefractive or semiconductor medium \citep{Lega_1994,HMT-PRL-95,SSW-PRL-97,MM-PRE-02}.
Diffraction control was also proposed using metamaterials \citep{GCdST-PRA-07,KTV-JOSAB-09,GGVdS-PRA-08,TKG-PRA-11}
or atomic resonances \citep{FLS-NAP-09}. In this paper, we study
how aberrations appear as leading effects in nearly-degenerate broad-area
surface-emitting lasers. We show that these effects can crucially
modify the spatio-temporal mode-locking dynamics and give rise to
either spatially tilted pulses or spatio-temporal solitons. Our results
represent a step further towards the understanding of spatio-temporal
mode-locking \citep{WSP-Nature-20,DXL-PRL-22,WRC-Optica-22}, the
realization of fully confined three-dimensional light bullets \citep{J-PRL-16,GRM-PRL-17,GJ-PRA-17}
and may lead to new ideas and applications for beam steering, tweezing
\citep{ADB-OL-86} and tailored optical energy potential landscapes
\citep{WAE-LPR-13,BCL-NAR-21}.

The dynamics of the transverse profile of the field close to the SIC
and in presence of spherical aberrations is equivalent to the dynamics
of a quantum harmonic oscillator featuring a fourth-order derivative
\begin{align}
-i\partial_{\theta}A & =\left(cx^{2}+b\partial_{x}^{2}+s\partial_{x}^{4}\right)A\,,\label{eq:funny_oscillator}
\end{align}
where $c$, $b$ and $s$ are real coefficients that correspond to
the residual wavefront curvature, diffraction, and spherical aberration,
respectively. We will focus our attention on spherical aberration
as it is often the dominant form of aberration. A detailed derivation
is provided in the Appendix I. Notice that Eq.~\eqref{eq:funny_oscillator}
was studied already albeit in a different context in \citep{SHV-PRE-06}. 

It is well known that the potential $V\left(x\right)=c\,x^{2}$ in
Eq.~\eqref{eq:funny_oscillator} supports localized solutions for
$bc<0$ \citep{SIEGMAN-BOOK}. Setting $A(x,\theta)=A_{{\rm s}}(x)e^{-i\omega\theta}$
with $\omega$ denoting the frequency of the eigenmode and imposing
boundedness of the the solution defines a singular Sturm-Liouville
problem (SLP) that allows determining $\omega$: 
\begin{equation}
\left(\omega+cx^{2}+b\partial_{x}^{2}+s\partial_{x}^{4}\right)A_{s}=0\,,\lim_{x\rightarrow\pm\infty}A_{s}=0\label{eq:SL1}
\end{equation}
For $bc<0$ and $s=0$ the solutions of Eq.~\eqref{eq:SL1} are the
so-called Hermite-Gauss (HG) modes $H_{n}(\frac{x}{\sigma})$ with
$\sigma=\sqrt{-b/c}$ and frequencies $\omega_{n}=\dfrac{b}{\sigma^{2}}(2n+1)$
while $n\in\mathbb{N}$ is the modal index. We will see later that
this situation corresponds to the case of a stable cavity devoid of
aberration. 

The HG modes are invariant under Fourier transform since the multiplicative
and differential terms in Eq.~\eqref{eq:funny_oscillator} are exchanged
upon performing the latter. However, the role played by the fourth
order derivative in Eq.~\eqref{eq:funny_oscillator} becomes more
clear in Fourier space, where the SLP becomes 
\begin{align}
\left(\omega+c\partial_{k}^{2}-bk^{2}+sk^{4}\right)\hat{A}_{s} & =0\,,\label{eq:SLP}
\end{align}
Here, we defined $\hat{A}_{s}$ as the Fourier transform of $A_{s}$.
Hence, the presence of a fourth order derivative in Eq.~Eq.~\eqref{eq:funny_oscillator}
converts the situation to that of a particle in single or a double-well
potential. In case of $b<0$ and $c>0$ the potentials $V\left(x\right)$
or $\hat{V}\left(k\right)=-b\,k^{2}+sk^{4}$ both feature a minimum
at the origin as shown in Fig.~\ref{fig1}~(a).

\begin{figure}
\includegraphics[viewport=347bp 0bp 627bp 242bp,clip,width=1\columnwidth]{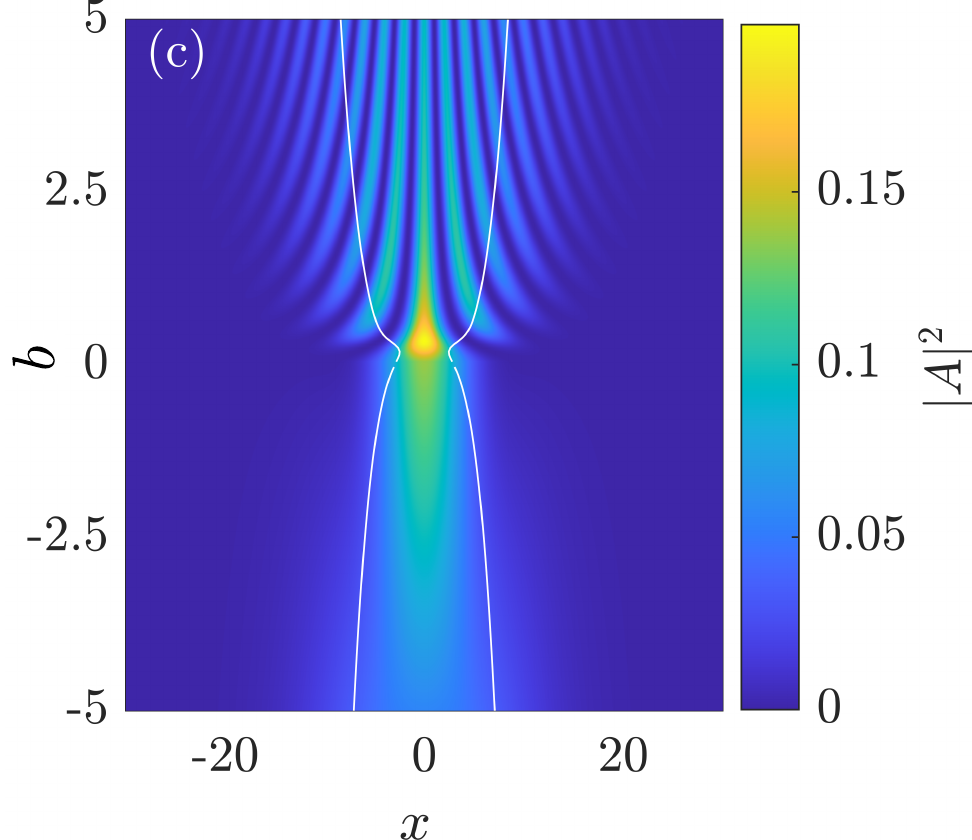}
\caption{Numerical solution of Eq.~\eqref{eq:funny_oscillator} for different
values of $b$ and fixed values of $s=1$ and $c=4.97\times10^{-4}$.
White line indicates the second moment.}
\label{fig1b}
\end{figure}

The bounded solutions of Eq.~\eqref{eq:funny_oscillator} remain
practically unaffected by small values of $s$, since the solutions
remain concentrated at low values of $k$ for which $bk^{2}\gg sk^{4}$.
Let us know consider the situation where $b$ changes its sign, which
usually signals the transition from a stable to an unstable cavity.
In that case, the potential $\hat{V}(k)$ develops a negative curvature
around $k=0$ for $b>0$. If there were no aberrations present in
the system, i.e. $s=0$, the laser would simply turn off as there
is no transverse mode to support emission. Mathematically speaking,
this means that the bounded solutions of the SLP cease to exists. 

However, the situation can be remedied by the fourth order derivative
as depicted in Fig.~\ref{fig1}(b): For $s>0$, two new minima emerge
symmetrically at $k_{0}=\pm\sqrt{b/2s}$ which allows bounded solutions
to continue to exist in the otherwise unstable domain. The ground
state solution $\hat{A}_{s}(k)$ in Fourier space can be approximated
by a superposition of two bell-shaped functions localized around $\pm k_{0}$
which, in real space, amounts to a strongly modulated eigenmode with
wavelength $\lambda_{\perp}=2\pi/k_{0}$, akin to the interference
between Bose-Einstein condensates in a double-well potential~\citep{Ketterle:Science:1997,Ketterle:PRL:2004}.
This situation is depicted in Fig.~\ref{fig1b} where we solved for
the ground state of Eq.~(\ref{eq:funny_oscillator}) numerically
using complex time evolution. Note, that the waist of the mode decreases
monotonically upon approaching $b=0$, but remains finite at $b=0$.
This is due to the fact that the higher order derivative start to
play a more dominant role. In the unstable range $b>0$ we notice
that the second order moment increases, albeit at at a faster rate.

\begin{figure*}[!t]
\includegraphics[width=1\columnwidth]{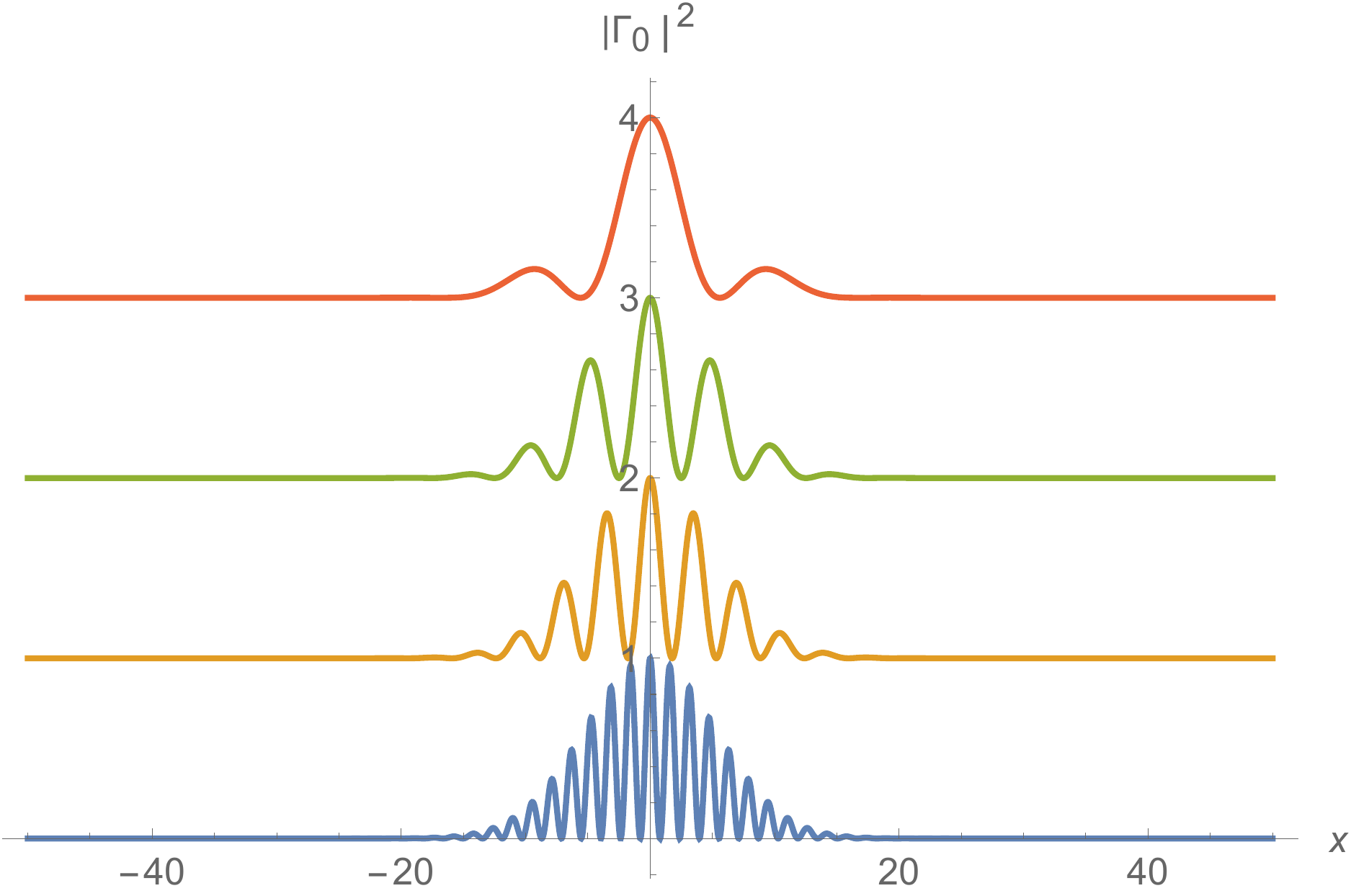}\includegraphics[width=1\columnwidth]{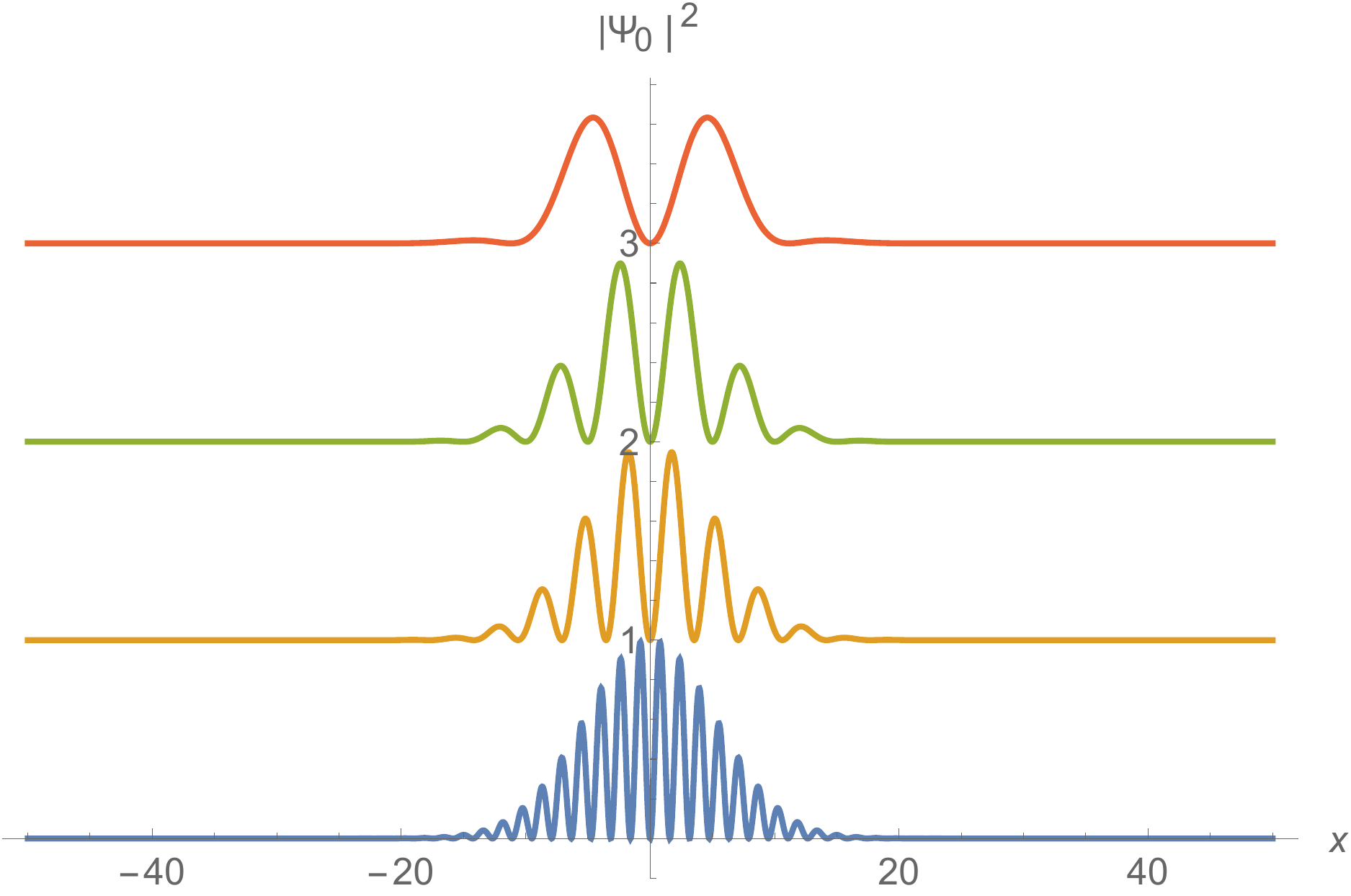}

\caption{Evolution of $\left|\Gamma_{0}\left(x,0\right)\right|^{2}$and $\left|\Psi_{0}\left(x,0\right)\right|^{2}$
as a function of $s$. Traces are shifted for clarity. From top to
bottom, $s=\left(5,\,1,\,0.5,\,0.1\right)$. Other parameters are
$\left(b,\,c\right)=\left(0.7873,\,0.0004973\right)$.}
\label{fig:Gamma_Psi_S}
\end{figure*}

\begin{figure*}[!t]
\includegraphics[width=1\columnwidth]{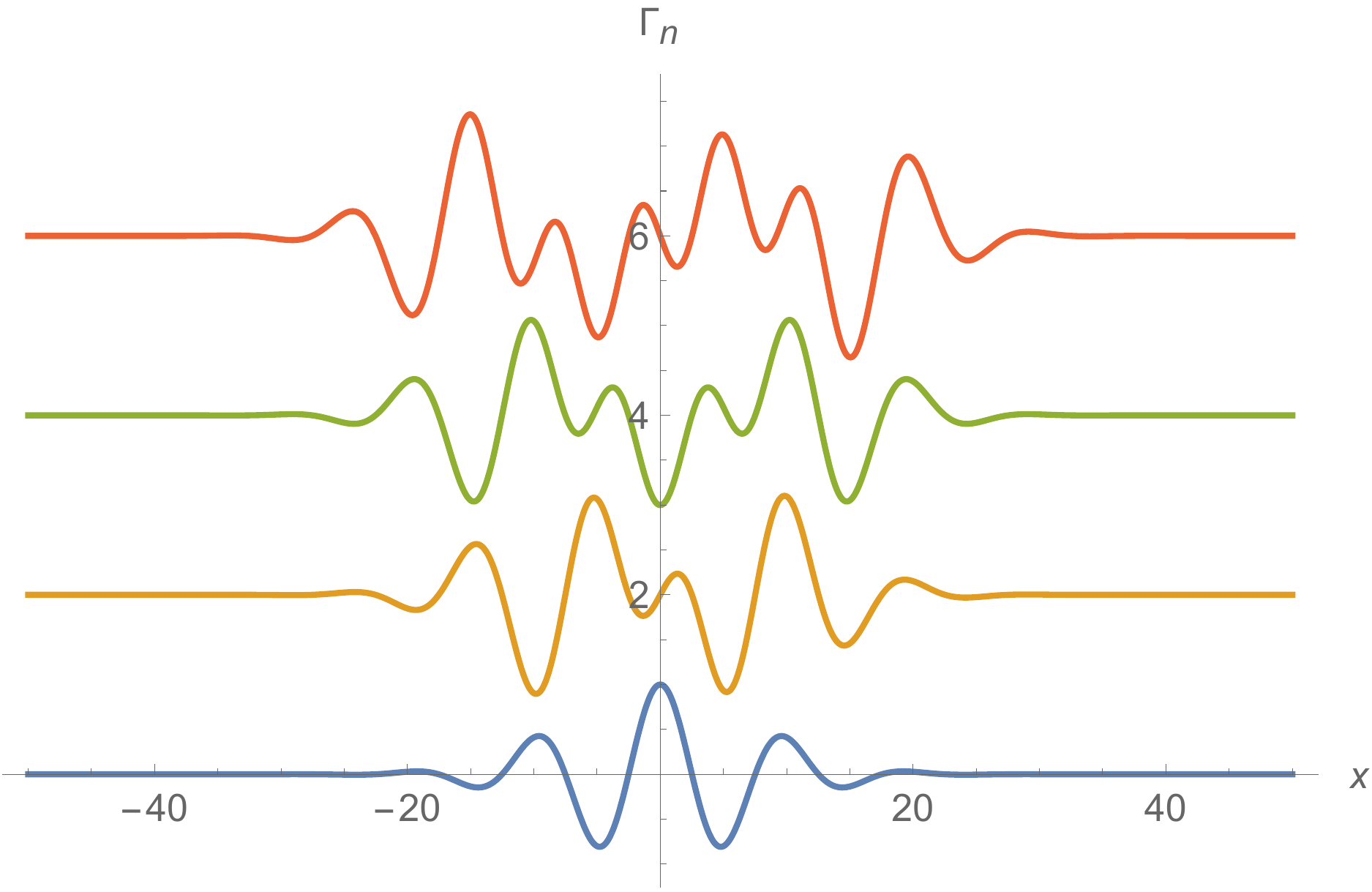}\includegraphics[width=1\columnwidth]{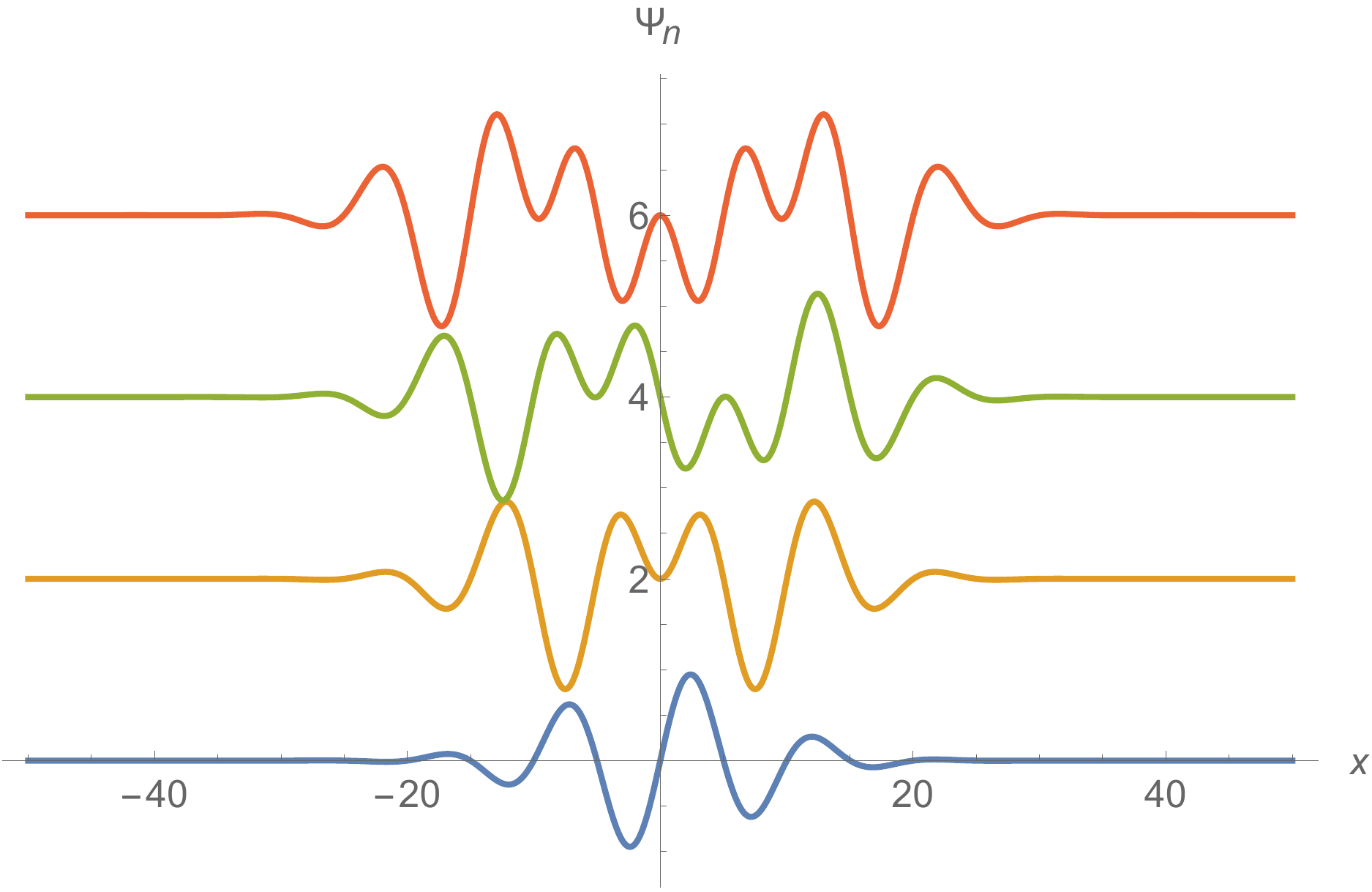}

\caption{Evolution of $\Gamma_{n}\left(x,0\right)$and $\Psi_{n}\left(x,0\right)$
as a function of the modal index $n$. We note the even/odd alternating
sequence. Traces are shifted for clarity. From top to bottom, $n=\left(3,\,2,\,1,\,0\right)$.
Other parameters are $\left(b,\,c,\,s\right)=\left(0.7873,\,0.0004973,1\right)$.}
\label{fig:Gamma_Psi_0123}
\end{figure*}

While we could not obtain closed form analytical solutions of Eq.~(\ref{eq:SL1})
for $s\neq0$, the following change of variable 
\begin{eqnarray}
A\left(x\right) & = & A_{s}\left(x\right)\exp\left[ik_{0}x-\omega\theta\right]\,,\;k_{0}=\sqrt{\frac{b}{2s}}
\end{eqnarray}
permits cancelling the first-order derivative and leads to 
\begin{equation}
\left(\omega-\frac{b^{2}}{4s}+cx^{2}-2b\partial_{x}^{2}+i\sqrt{8bs}\partial_{x}^{3}+s\partial_{x}^{4}\right)A_{s}=0\,.\label{eq:HGU}
\end{equation}
Slowly varying envelope solutions, where the characteristic length
scale for the envelope $A_{s}$ is much larger than $2\pi/k_{0}$
correspond to the inequality $\left|\partial_{x}A_{s}\right|\ll\left|k_{0}A_{s}\right|$.
If this condition is fulfilled, one may neglect the third and fourth
order derivative in Eq.~\eqref{eq:HGU}, leading to 
\begin{eqnarray}
\left(\omega-\frac{b^{2}}{4s}+cx^{2}-2b\partial_{x}^{2}\right)A_{s} & = & 0\,.\label{eq:unstableHG}
\end{eqnarray}
By comparing Eq.~\eqref{eq:funny_oscillator} with Eq.~\eqref{eq:SL1}
we note that the latter is again a Hermite-Gauss equation where the
parameter $b$ has been replaced by $-2b$, which fully explains the
increased diverging rate of the second moment in Fig.~\ref{fig1b}.
We note that the frequencies $\omega$ are shifted due to the additional
term $b^{2}/\left(4s\right)$. 

In summary, we disclose the surprising result that an unstable cavity
does not necessarily turn off upon crossing the SIC, and lasing eigenmodes
can still be supported by spherical aberrations. The modes can be
approximated by 
\begin{eqnarray}
A\left(x,\theta\right) & = & H_{n}\left(\frac{x}{\sigma}\right)\exp\left[i\left(k_{0}x-\omega_{n}\theta\right)\right]\,,\\
\omega_{n} & = & \frac{b^{2}}{4s}-\frac{2b}{\sigma^{2}}\left(2n+1\right)\,,\\
\sigma^{2} & = & \sqrt{\frac{2b}{c}}\,,\,k_{0}=\sqrt{\frac{b}{2s}}\,.
\end{eqnarray}

\begin{figure*}[!t]
\begin{centering}
\includegraphics[width=2.1\columnwidth]{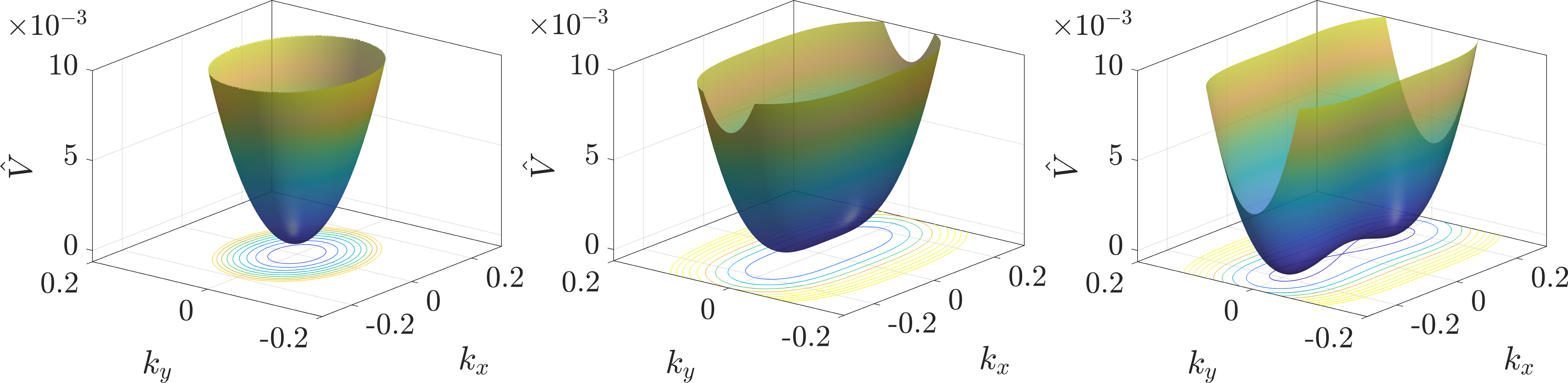}
\par\end{centering}
\caption{Birth of two symmetrically located minima for the Fourier potential
$\hat{V}$ for a stable, marginal and unstable cavity with (from left
to right) $b_{x}=\left(-0.2,0,0.05\right)$ Other parameters are $\left(b_{y},s\right)=\left(-0.8,1\right)$.}
\label{fig:2D_disp_rel}
\end{figure*}

Since $k_{0}\in\mathbb{R}$, it is convenient to define two families
of eigenfunctions $\Gamma_{n}\left(x,\theta\right)$ and $\Psi_{n}\left(x,\theta\right)$
as: 
\begin{eqnarray}
\Gamma_{n}\left(x,\theta\right) & = & H_{n}\left(\frac{x}{\sigma}\right)\cos\left(k_{0}x\right)e^{-i\omega_{n}\theta}\,,\label{eq:GauCosine}\\
\Psi_{n}\left(x,\theta\right) & = & H_{n}\left(\frac{x}{\sigma}\right)\sin\left(k_{0}x\right)e^{-i\omega_{n}\theta}.\label{eq:GauSine}
\end{eqnarray}
The latter are depicted in Fig.~\ref{fig:Gamma_Psi_S} and Fig.~\ref{fig:Gamma_Psi_0123},
respectively. We observe the effect of the parameter $s$ on the ground
states $\Gamma_{0}$ and $\Psi_{0}$ in Fig.~\ref{fig:Gamma_Psi_S}.
In order to stabilize a bounded eigenmode, the decrease of $s$ must
be compensated by additional oscillations and, therefore, an increase
in $k_{0}\sim1/\sqrt{s}$. In Fig.~\ref{fig:Gamma_Psi_0123} we also
depict how the intensity of the modes $\Gamma_{n}\left(x,\theta\right)$
and $\Psi_{n}\left(x,\theta\right)$ alternate between even and odd
when varying the modal index $n$. 

We now turn our attention to the two-dimensional (2D) transverse profile
whose evolution is governed by
\begin{eqnarray}
\negthickspace\negthickspace\negthickspace\negthickspace\negthickspace-i\frac{\partial A}{\partial\theta} & = & \left(c_{x}x^{2}+c_{y}y^{2}+b_{x}\partial_{x}^{2}+b_{y}\partial_{y}^{2}+s\nabla_{\perp}^{4}\right)A.\label{eq:A2D}
\end{eqnarray}
Here, we assume that the cavity possesses two orthogonal axes denoted
$\boldsymbol{r}_{\perp}=\left(x,y\right)$ and, for the sake of generality,
we introduced dichroism in Eq.~\ref{eq:A2D}, i.e. $b_{x}\neq b_{y}$
and $c_{x}\neq c_{y}$. The effect of spherical aberration in 2D translates
into a bilaplacian operator $\nabla_{\perp}^{4}=\partial_{x}^{4}+2\partial_{x}^{2}\partial_{y}^{2}+\partial_{y}^{4}$.
We note that $\nabla_{\perp}^{4}$ is the only fourth order operator
that preserves the rotational symmetry both in real and in Fourier
space, which is consistent with the idea of spherical. The equivalent
SLP for $A\left(\boldsymbol{r}_{\perp},\theta\right)=A_{s}\left(\boldsymbol{r}_{\perp}\right)\exp\left(-i\omega\theta\right)$
reads 
\begin{eqnarray}
\negthickspace\negthickspace\negthickspace\negthickspace\negthickspace0 & = & \left(\omega+c_{x}x^{2}+c_{y}y^{2}+b_{x}\partial_{x}^{2}+b_{y}\partial_{y}^{2}+s\nabla_{\perp}^{4}\right)A_{s}.\label{eq:SLP_2D}
\end{eqnarray}
When $s=0$ and the two conditions $b_{x}c_{x}<0$ and $b_{y}c_{y}<0$
are simultaneously verified, the SLP admits separable solutions that
are simply the product of the HG modes in the $x$ and $y$ directions
discussed previously \citep{SIEGMAN-BOOK}. We now consider the effect
of aberration close to the SIC when the cavity becomes unstable. Due
to the inherent dichroism present in any realistic experimental system,
one can assume that the cavity becomes unstable first in one direction,
say the $x$-direction. Applying the Fourier transform to Eq.~\ref{eq:SLP_2D}
leads to 
\begin{eqnarray}
0 & = & \left[\omega+c_{x}\partial_{q_{x}}^{2}+c_{y}\partial_{q_{y}}^{2}+\hat{V}\left(\boldsymbol{k}_{\perp}\right)\right]A_{s},\label{eq:SLP_2DF}
\end{eqnarray}
where we defined the potential in Fourier space
\begin{eqnarray}
\hat{V}\left(\boldsymbol{k}_{\perp}\right) & = & -b_{x}k_{x}^{2}-b_{y}k_{y}^{2}+s\left(k_{x}^{2}+k_{y}^{2}\right)^{2}\,.
\end{eqnarray}
A representation of $\hat{V}\left(\boldsymbol{k}_{\perp}\right)$
for three different values of $b_{x}$ is given in Fig.~\ref{fig:2D_disp_rel}.
We see that the transition from a stable to an unstable cavity is
obtained when $b_{x}$ changes its sign from negative to positive
which gives rise to the appearance of two new minima as observed in
the 1D case in Fig.~\ref{fig1}. Their position is obtained by simply
setting $\partial_{k_{x}}\hat{V}=\partial_{k_{y}}\hat{V}=0$ which
leads to $k_{\perp}^{\pm}=\pm\left(\sqrt{\frac{b_{x}}{2s}},0\right)$.
As already discussed, this corresponds to off-axis emission on the
$x$-axis and stable on-axis emission on the $y$-axis. The curvature
of $\hat{V}\left(k_{\perp}\right)$ around the two minima in $k_{\perp}^{\pm}$
upon back-transforming to direct space corresponds to the effective
diffraction experienced by a wave-packet centered around this tilted
wavevector $k_{\perp}^{\pm}$. We obtain
\begin{eqnarray}
\frac{1}{2}\frac{\partial^{2}\hat{V}}{\partial k_{x}^{2}} & = & 2b_{x}\;,\;\frac{1}{2}\frac{\partial^{2}\hat{V}}{\partial k_{y}^{2}}=b_{x}-b_{y}\;,\;\frac{\partial^{2}\hat{V}}{\partial k_{x}\partial k_{y}}=0\,.\label{eq:Curvature}
\end{eqnarray}
The curvature in the $x$-direction changed from $-b_{x}\rightarrow2b_{x}$,
similarly to the 1D case. However, the perpendicular direction is
also affected by the off-axis emission, leading to the substitution
$b_{y}\rightarrow b_{y}-b_{x}$. In the case of a cavity that changes
behaviour from stable $b_{x}<0$ to unstable $b_{x}>0$, the off-axis
emission along the $x$-direction that re-stabilizes the emission
renders the perpendicular direction ``more diffractive'' since $b_{y}-b_{x}<b_{y}$
if $b_{x}>0$. The results of Fig.~\ref{fig:2D_disp_rel} entice
us to seek again a modulated profile 
\begin{eqnarray}
A\left(\boldsymbol{r}_{\perp}\right) & = & A_{s}\left(\boldsymbol{r}_{\perp}\right)\exp\left[i\left(k_{0}x-\omega\theta\right)\right]
\end{eqnarray}
with $k_{0}=\sqrt{b_{x}/\left(2s\right)}$.  Truncating to second
order we get the approximate SLP problem
\begin{eqnarray}
\negthickspace\negthickspace\negthickspace\negthickspace\negthickspace\frac{b_{x}^{2}}{4s}-\omega & = & \left[c_{x}x^{2}+c_{y}y^{2}-2b_{x}\partial_{x}^{2}+\left(b_{y}-b_{x}\right)\partial_{y}^{2}\right]A_{s},\label{eq:2Dmod}
\end{eqnarray}
where the modification of the diffraction coefficients in Eq.~\ref{eq:2Dmod}
is fully consistent with the discussion of the curvature of $\hat{V}$
given in Eq.~\ref{eq:Curvature}. The eigenmodes of such an unstable
cavity are the product of a modulated HG mode in the unstable direction
and of a regular HG mode in the y-direction. As such, we define the
eigenmode family with two modal indices $\left(n,m\right)$ 
\begin{eqnarray}
\Gamma_{n,m}\left(\boldsymbol{r}_{\perp},\theta\right) & = & H_{n}\left(\frac{x}{\sigma_{x}}\right)H_{m}\left(\frac{y}{\sigma_{y}}\right)\cos\left(k_{0}x\right)e^{-i\omega_{n,m}\theta}\,,\nonumber \\
\Psi_{n,m}\left(\boldsymbol{r}_{\perp},\theta\right) & = & H_{n}\left(\frac{x}{\sigma_{x}}\right)H_{m}\left(\frac{y}{\sigma_{y}}\right)\sin\left(k_{0}x\right)e^{-i\omega_{n,m}\theta}\,,\nonumber \\
\omega_{n,m} & = & \frac{b_{x}^{2}}{4s}-\frac{2b_{x}}{\sigma_{x}^{2}}\left(2n+1\right)+\dfrac{b_{y}}{\sigma_{y}^{2}}(2m+1)\,,\label{eq:HGM_2D}\\
\sigma_{x} & = & \sqrt{2b_{x}/c_{x}}\,,\nonumber \\
\sigma_{y} & = & \sqrt{\left(b_{x}-b_{y}\right)/c_{y}}\,.\nonumber 
\end{eqnarray}

\begin{figure}
\includegraphics[width=1\columnwidth]{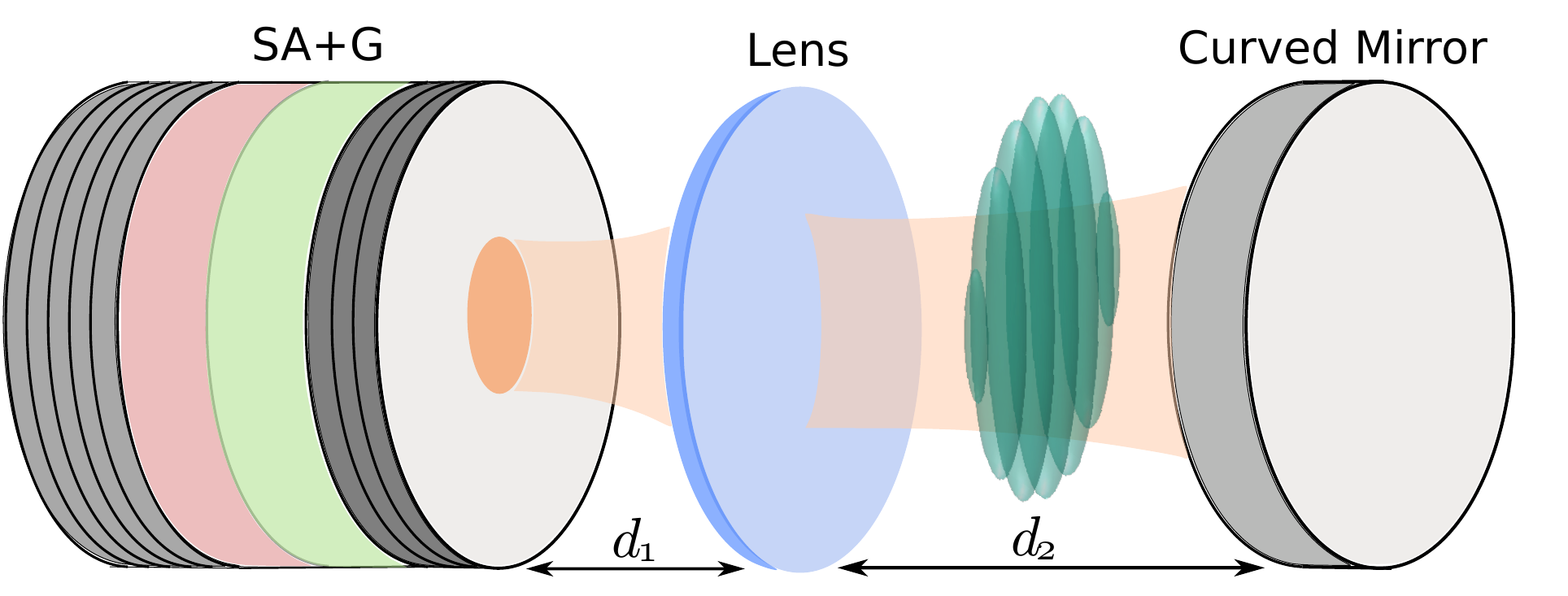}
\caption{A schematic of the MIXSEL, where both the gain (green) and the saturable
absorption (pink) are contained in the same micro-cavity. It is coupled
face-to-face to a distant external mirror by an imperfect self-imaging
system.}
\label{fig:mixsel}
\end{figure}

The situation considered in Eq.~\eqref{eq:funny_oscillator} materializes
for instance in the study of passively mode-locked integrated external-cavity
surface-emitting laser (MIXSELs) in a near-degenerate cavity~\citep{HBE-OE-08}
as depicted in Fig.~\ref{fig:mixsel}. Here, the gain (G) and saturable
absorber (SA) media are enclosed in a single micro-cavity and the
external mirror is assumed to be ideal while the collimating lens
is not. Our modeling approach is based on a Haus master equation model
for passive mode-locking (PML) adapted to the experimentally relevant
long cavity regime regime~\citep{CJM-PRA-16,GJ-PRA-17,SHJ-PRAp-20,Hausen2020a},
where the PML pulses become individually addressable temporal localized
states (TLSs). The Haus equation relates the slow evolution of the
three-dimensional intra-cavity field $E\left(\boldsymbol{r}_{\perp},t,\theta\right)$
to the dynamics of the population inversion in the gain $N_{1}\left(\boldsymbol{r}_{\perp},t\right)$
and the saturable absorber $N_{2}\left(\boldsymbol{r}_{\perp},t\right)$
as 
\begin{eqnarray}
\negthickspace\negthickspace\negthickspace\negthickspace\negthickspace\frac{\partial E}{\partial\theta} & \negthickspace=\negthickspace & \left[\left(1-i\alpha_{1}\right)N_{1}+\left(1-i\alpha_{2}\right)N_{2}-\kappa+\mathcal{L}\right]E,\label{eq:HausE}\\
\negthickspace\negthickspace\negthickspace\negthickspace\negthickspace\frac{\partial N_{1}}{\partial t} & \negthickspace=\negthickspace & \gamma_{1}\left(J_{1}-N_{1}\right)-N_{1}\left|E\right|^{2},\,N_{1}(\boldsymbol{r_{\perp}},0)=J_{1},\label{eq:HausN1}\\
\negthickspace\negthickspace\negthickspace\negthickspace\negthickspace\frac{\partial N_{2}}{\partial t} & \negthickspace=\negthickspace & \gamma_{2}\left(J_{2}-N_{2}\right)-\hat{s}N_{2}\left|E\right|^{2}\,,N_{2}(\boldsymbol{r}_{\perp},0)=J_{2}.\label{eq:HausN2}
\end{eqnarray}

In this formalism, the variable $t\in\left[0,\tau\right]$ represents
the round-trip time in the external cavity and $\theta$ is a second
dimensionless time scale normalized by $\tau$. The latter corresponds
to the slow evolution of the pulse under the combined effect of gain,
absorption and spatio-temporal filtering. The two transverse dimensions
are denoted $\boldsymbol{r}_{\perp}=\left(x,y\right)$ and allow for
pattern formation in the plane perpendicular to the propagation direction
of the pulse, cf. Fig.~\ref{fig:mixsel}. Note that in the long cavity
regime, the spatio-temporal distributions of the carriers $N_{j}\left(\boldsymbol{r}_{\perp},t\right)$
are slaved to the evolution of the optical field \citep{CJM-PRA-16,Hausen2020a}
and do not depend explicitly on the slow timescale $\theta$. In this
regime, one can safely assume a full recovery of the carrier between
pulses and a total loss of memory from one round-trip toward the next
one. We define in Eqs.~(\ref{eq:HausE})-(\ref{eq:HausN2}) $\kappa$
as the round-trip cavity losses, and $\alpha_{j}$ are the linewidth
enhancement factors of the two active media that relax with timescales
$\gamma_{j}^{-1}$ toward the equilibrium values $J_{j}$. The ratio
of the saturation fluence of the gain and of the SA is denoted by
$\hat{s}$. A standard speudo-spectral split-step method was used
for the numerical simulations \citep{GJ-PRA-17}. The effective cavity
\emph{spatio-temporal linear operator} $\mathcal{L}$, accounts for
the finite gain bandwidth and chromatic dispersion, but also for non-perfect
imaging conditions, diffraction, parabolic wavefront curvature and
mirror losses due to finite aperture. $\mathcal{L}$ is given by
\begin{eqnarray}
\mathcal{L} & = & d_{g}\partial_{t}^{2}+\mathcal{L}_{\perp}\,,\label{eq:Lin_op}
\end{eqnarray}
where $d_{g}$ is the temporal diffusion coefficient representing
the gain bandwidth. 

The Fresnel transform~\citep{PB-JOSAA-97} permits calculating the
transverse effects occurring at each round-trip analytically from
the round-trip (ABCD) matrix in the paraxial approximation \citep{SIEGMAN-BOOK}.
In presence of aberrations, its calculation is achieved by expanding
the exact (non-paraxial) operator corresponding to the lens as a sum
of an ideal element plus a deviation. The latter potentially contains
all the wavefront curvature contributions beyond the parabolic approximation.
Assuming a large ratio between the focal length of the mirror and
that of the lens permits expressing the spherical aberrations as a
bilaplacian operator, see the Appendix I for details. The transverse
round-trip operator $\mathcal{L}_{\perp}$ is given by 
\begin{eqnarray}
\mathcal{L}_{\perp} & = & icx^{2}+\left(d_{f}+ib\right)\partial_{x}^{2}+is\partial_{x}^{4}\,.\label{eq:Lin_top}
\end{eqnarray}
The finite size of lenses and the numerical aperture of the whole
optical system is modeled by as a soft aperture and a real transverse
diffusion parameter $d_{f}$. Finally, $b$ is the normalized paraxial
diffraction parameter, $c$ is the normalized parabolic wavefront
curvature that are the off-diagonal elements of the round-trip ABCD
matrix \citep{SIEGMAN-BOOK} and $s$ is the spherical aberration
parameter. We note that moving the lens in Fig.~\ref{fig:mixsel}
modifies $c$ while moving the mirror modifies both $b$ and $c$. 

A qualitative model for the transverse profile of the TLSs such as
the one derived in \citep{J-PRL-16,GJ-PRA-17} can be obtained by
essentially adapting New's method for PML \citep{N-JQE-74} to the
situation at hand. This method exploits the scale separation occurring
between the pulse evolution, the so-called fast stage in which stimulated
emission is dominant, and the slow stage that is controlled by the
gain recovery processes. Assuming, as in \citep{J-PRL-16,GJ-PRA-17},
that the four-dimensional spatio-temporal profile $E\left(x,t,\theta\right)$
can be factored as $E\left(x,t,\theta\right)=A\left(x,\theta\right)p\left(t\right)$
with $p\left(t\right)$ a normalized TLS profile, one obtains a Rosanov
equation \citep{RK-OS-88} for the slow evolution of the TLS transverse
profile $A\left(x,\theta\right)$ as 
\begin{eqnarray}
\negthickspace\negthickspace\partial_{\theta}A & \negthickspace= & \left[f\left(\left|A\right|^{2}\right)+icx^{2}+\left(d_{f}+ib\right)\partial_{x}^{2}+is\partial_{x}^{4}\right]A.\label{eq:Rosa1}
\end{eqnarray}
The normalized coefficients $b$, $c$ and $s$ represent the residual
paraxial diffraction, parabolic wavefront curvature and aberration
coefficients close to SIC. Their expression and their derivation are
given in the Appendix I. We define the effective nonlinearity as 
\begin{equation}
f\left(P\right)=\left(1-i\alpha_{1}\right)J_{1}\,g\left(P\right)+\left(1-i\alpha_{2}\right)J_{2}\,g\left(\hat{s}P\right)-\kappa\,,\label{eq:Rosa2}
\end{equation}
with $J_{1,2}$ and $\alpha_{1,2}$ denoting the bias and the line-width
enhancement factors of the gain and absorber, $\hat{s}$ the ratio
of their saturation energies and $\kappa$ the round-trip losses.
The nonlinear response of the active material to a pulse is given
by $g\left(P\right)=\left(1-e^{-P}\right)/P$ \citep{N-JQE-74,J-PRL-16,GJ-PRA-17}.
\begin{figure}[h]
\includegraphics[width=0.98\columnwidth]{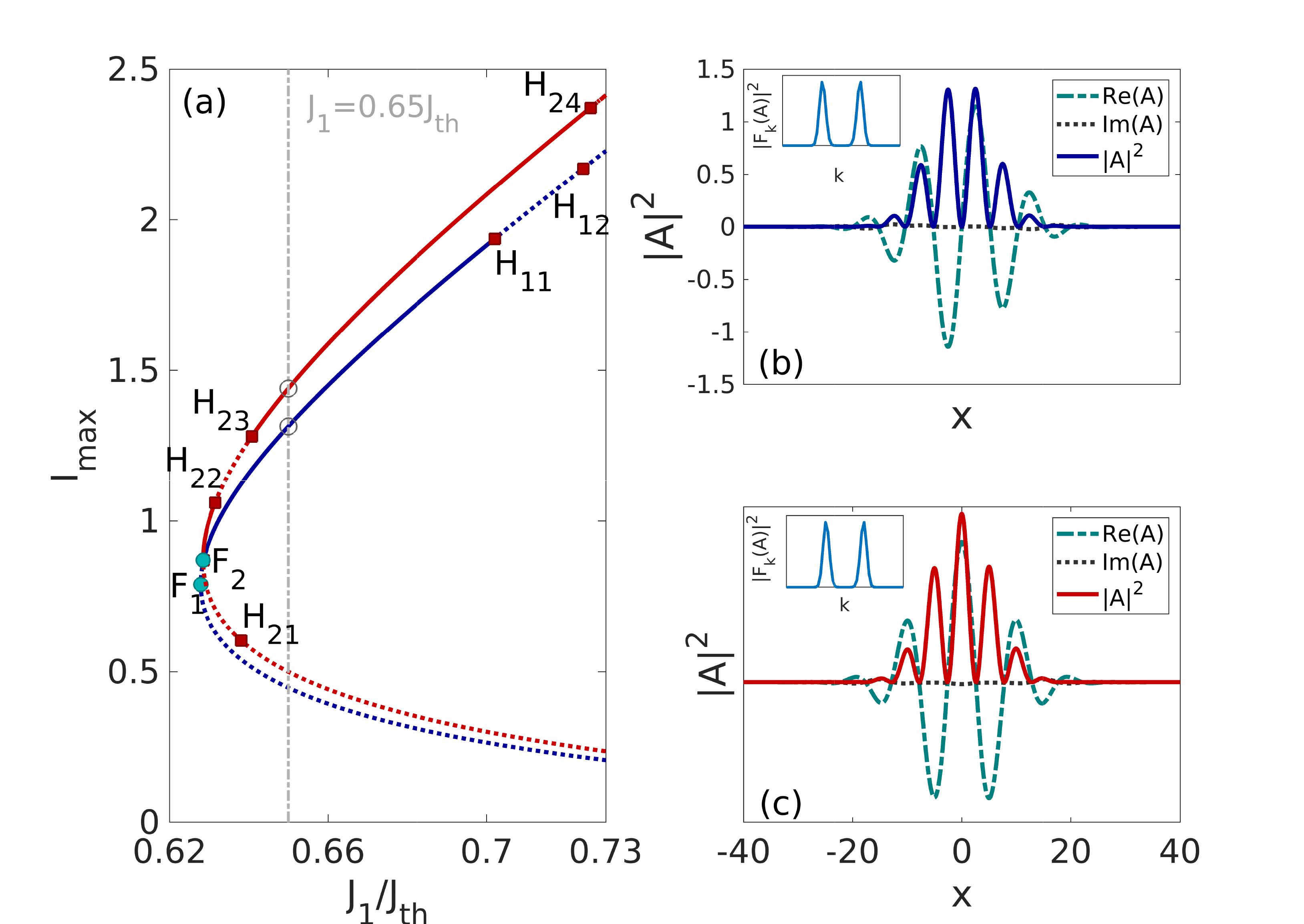}
\caption{(a) Branches of one-dimensional tilted HG modes $\Gamma_{0}$ (red)
and $\Psi_{0}$ (blue) of Eq.~\eqref{eq:Rosa1} as a function of
the normalized gain bias $J_{1}/J_{\mathrm{th}}$. The mode $\Gamma_{0}$
is stable (solid line) between the fold $F_{2}$ (cyan circle) and
the AH point $H_{22}$ as well as between AH points $H_{23}$ and
$H_{24}$ (red squares), respectively. The mode $\Psi_{0}$ gains
the stability at the fold $F_{1}$ and remains stable up to the AH
bifurcation point $H_{11}$. (b,c) displays exemplary profiles of
$\Gamma_{0}$ and $\Psi_{0}$ for $J_{1}=0.65J_{\mathrm{th}}$ (gray
dashed line). Real (turkis, dashed), imaginary (black, dotted) and
intensity fields (blue, red, solid), respectively, are presented.
Parameters are $\alpha_{1}=1.5,\alpha_{2}=0.5,J_{2}=-0.06,\hat{s}=15,\kappa=0.035,d_{f}=10^{-4},b=0.78,c=4.97\cdot10^{-4},s=1.0$.}
\label{fig:branch_sinecos_J1} 
\end{figure}

The equations~(\ref{eq:Rosa1},\ref{eq:Rosa2}) provide a unified
framework which allow bridging our results for spatio-temporal dynamics
with the former results of \citep{RK-OS-88,VFK-JOB-99} for the case
of static auto-solitons in bistable interferometers. There, the function
$g\left(P\right)$ should be replaced by the saturated lineshape transition
$\sim\left(1+P\right)^{-1}$. As such, our discussion of the effect
of aberrations close to SIC is equally valid for temporal solitons,
regular mode-locking and CW beams.

In one-dimension, Eq.~\eqref{eq:funny_oscillator} is recovered for
the empty cavity, i.e., $f=0$. Hence, we expect the emergence of
a \emph{stable} family of $\Gamma_{n}$ and $\Psi_{n}$ tilted HG
modes (\ref{eq:GauCosine},\ref{eq:GauSine}) in the \emph{unstable
cavity}, where the condition $bc<0$ is violated. We performed a bifurcation
analysis of Eq.~(\ref{eq:Rosa1}) in one transverse dimension using
path-continuation framework pde2path~\citep{uecker2014}. The results
are summarized in Fig.~\ref{fig:branch_sinecos_J1} in the unstable
cavity regime, where the peak powers for the fundamental tilted modes
$\Gamma_{0}$ (red) and $\Psi_{0}$ (blue) are shown in the panel
(a) as a function of the gain bias $J_{1}$ normalized to the threshold
value $J_{th}$. Furthermore, two exemplary profiles of $\Gamma_{0}$
and $\Psi_{0}$ at the same fixed gain value (gray dashed line) are
depicted in Fig.~\ref{fig:branch_sinecos_J1}~(b,c), respectively.
For both modes, we observe the typical subcritical transition that
leads to bistable TLSs as detailed in \citep{MJB-PRL-14,GJ-PRA-17};
The high intensity branch is stable while the lower branch is unstable
and creates a separatrix with the stable off solution. For the parameters
chosen, the leftmost limiting fold bifurcations $F_{1}$ and $F_{2}$
are almost identical for $\Gamma_{0}$ and $\Psi_{0}$. A small region
of the Andronov-Hopf (AH) instability for $\Gamma_{0}$ exists between
points $H_{22}$ and $H_{23}$. In this region, a small amplitude
oscillation is visible in time simulations. Both modes are limited
by the AH bifurcations $H_{11}$ and $H_{24}$, respectively, for
high gain values. We stress that these nonlinear modulated HG solutions
that are solutions of Eqs.~(\ref{eq:Rosa1},\ref{eq:Rosa2}), correspond
to a train of TLSs whose profile is a tilted beam supported by spherical
aberrations in an unstable cavity. 

\begin{figure*}
\centering \includegraphics[width=2\columnwidth]{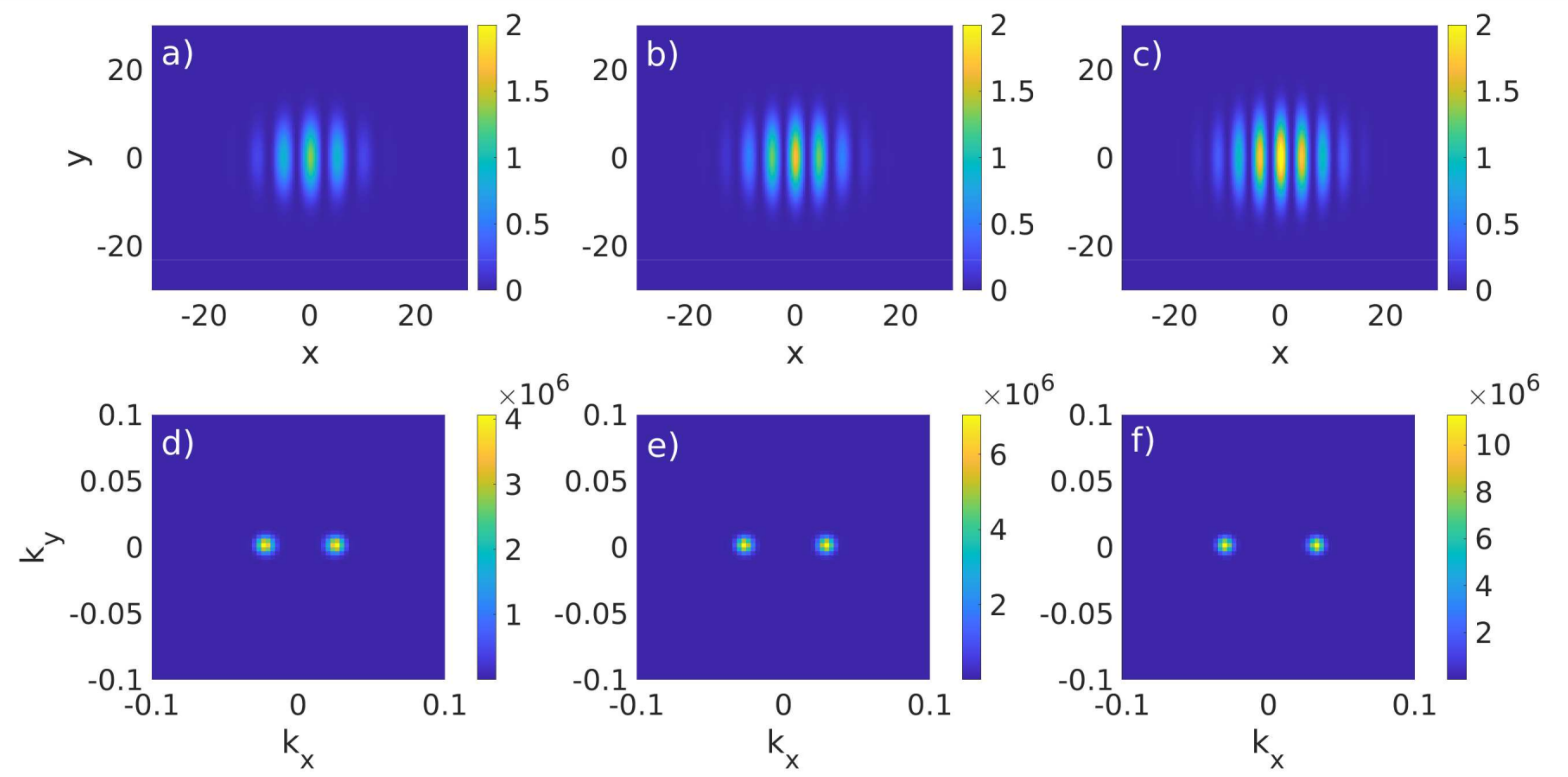}
\caption{(a-c): Intensity profiles of the 2D pattern obtained by the numerical
simulations for three values of $b_{x}=(0.78,\,1.02,\,1.25)$, respectively
at the fixed value of the current $J_{1}=0.65J_{\mathrm{th}}$. (d-f):
Corresponding power spectra in the $(k_{x},k_{y})$ plane. Parameters
are: $b_{y}=-0.39,c_{x}=c_{y}=4.97\cdot10^{-4}$. Other parameters
as in Fig.~\ref{fig:branch_sinecos_J1}.}
\label{fig:Optical_tiger}
\end{figure*}

In two spatial dimensions, we assume the system to be weakly astigmatic
and, as discussed previously, that the self-imaging condition is not
reached simultaneously for both transverse dimensions. For a cavity
that is stable in the vertical and unstable in the horizontal direction
we observed the tilted localized patterns shown in Fig.~\ref{fig:Optical_tiger}~(a-c)
by solving Eqs.~(\ref{eq:Rosa1},\ref{eq:Rosa2}) numerically. The
typical evolution for large values of $b_{x}$ (i.e., entering more
deeply into the unstable region) is presented together with the corresponding
far-field power spectrum $|\hat{A}|^{2}$. The panels (d-f) shows
the two peaks in the far field related to the value of the transverse
horizontal wave-number $\pm k_{0,x}$. Clearly, an increase of $b_{x}$
leads to an increase of $k_{0}$, which corresponds to a higher frequency
of the mode oscillations in the near-field. These results would correspond
closely to the experimental situation described either in \citep{BVM-XXX-24}
for a mode-locked vertical surface-emitting external-cavity semiconductor
laser using a saturable absorber or in \citep{NathanPHD} for a CW
broad area laser.

In conclusion, we have investigated the effect of wavefront aberrations
in a degenerated cavity. We found that the interplay between spherical
aberrations and the proximity to the SIC may lead to modulated beams
that can support either CW or temporal solitons in a mode-locked broad-area
VECSELs. These modulated beams are analogous to the eigenmodes of
a quantum particle in a double well potential. They can be analytically
approximated by spatially modulated Hermite-Gauss modes. We linked
the wavelength of their modulation to the parameters of the cavity.
We note that the transition from a stable towards an unstable cavity
around SIC can be obtained in two ways: Either for $c>0$ and changing
the sign of $b$ from negative to positive, or for $c<0$ and changing
the sign of $b$ correspondingly. These two situations are not identical
with respect to the effect of spherical aberrations since the off-axis
wavevector $k_{0}=\sqrt{b/2s}$ requires $b$ and $s$ to have the
same sign. As such, the sign of $s$ dictates the situation in which
one can observe these modulated Hermite-Gauss beams. While we focused
on the influence of spherical aberrations, we proposed a general framework
that, in principle, permits calculating the effect of the other Seidel
aberrations such as coma, distortion or field curvature in a cavity
close to SIC. We believe that exhibiting the link between these wavefront
curvature defects and their equivalent representations as spatial
operators provides a new framework that could lead to a range of interesting
new research avenues in photonics. Furthermore, the condition of a
large ratio between the focal distances of the aberrated lens and
the non-aberrated other elements could be relaxed. In this situation,
spherical aberrations translate into a non-local spatial operators
that may lead to rich pattern formation scenarios, as observed in
other fields \citep{FKK-PRL-03,MPS-PRL-16,JMG-PRL-17}.
\begin{acknowledgments}
We thank A.~Garnache M.~Guidici and M.~Marconi for useful discussions
regarding aberrations and self-imaging conditions and for their comments
on this manuscript. F.M and J.J acknowledge funding from the Ministerio
de Economía y Competitividad (PID2021-128910NB-100 AEI/FEDER UE, PGC2018-099637-B-100
AEI/FEDER UE). 
\end{acknowledgments}

\section*{Appendix I: Spherical aberration close to the SIC}

In this appendix we derive how the effect of spherical aberration
can be recast into the form of a bilaplacian differential operator
in the paraxial equation governing the field evolution after each
round-trip. We will use the formalism of wave optics provided by generalized
Huygens--Fresnel transform (HFT) for ABCD systems \citep{SIEGMAN-BOOK}.
We note that the HFT allows for the composition of parabolic operators,
i.e. quadratic wavefront profiles induced by parabolic lenses as well
as paraxial diffraction, since the latter also corresponds to a parabolic
operator in Fourier space. We characterize our system by an $\mathrm{ABCD}$
round-trip matrix 
\begin{eqnarray}
W & = & \left(\begin{array}{cc}
A & B\\
C & D
\end{array}\right)
\end{eqnarray}
with $\mathrm{det}\left(W\right)=AD-BC=1$. The generalized HFT for
the passage of light through a first-order optical system~\citep{PB-JOSAA-97}
composed of parabolic elements is given in one transverse dimension
by 
\begin{align}
O\left(x,l\right)= & \sqrt{\frac{-i}{\lambda B}}e^{ikl}\int_{-\infty}^{\infty}O\left(\xi,0\right)\times\label{eq:HF_ori}\\
 & \exp\left[i\frac{\pi}{\lambda B}\left(A\xi^{2}-2x\xi+Dx^{2}\right)\right]d\xi\,,\nonumber 
\end{align}
where $O\left(x,0\right)$ is the incoming field passing through the
system characterized by the $\mathrm{ABCD}$ matrix $W$ and $k=2\pi/\lambda$
is the wave-vector of light. We note that $B$ plays the role of a
propagation distance while $C$ is an inverse of a distance and represents
wavefront curvature. Further, $A,D$ are dimensionless quantities
that correspond to spatial and angular magnification, respectively. 

Close to the self-imaging condition we consider the limit $B\rightarrow0$
which renders Eq.~\eqref{eq:HF_ori} singular. This difficulty can
be avoided by factoring the quadratic form as follows: 
\begin{equation}
A\xi^{2}-2x\xi+Dx^{2}=A\left(\xi-\frac{x}{A}\right)^{2}+\left(D-\frac{1}{A}\right)x^{2}.
\end{equation}

Using that $\left(D-\frac{1}{A}\right)/B=C/A$, the Huygens-Fresnel
integral given in Eq.~\eqref{eq:HF_ori} becomes 
\begin{align}
O\left(x,l\right)= & e^{ikl}e^{i\frac{\pi}{\lambda}\frac{C}{A}x^{2}}\int_{-\infty}^{\infty}O\left(\xi,0\right)\times\label{eq:HFInt}\\
 & \sqrt{\frac{-i}{\lambda B}}\exp\left[i\frac{\pi A}{\lambda B}\left(\xi-\frac{x}{A}\right)^{2}\right]d\xi\,.\nonumber 
\end{align}
The equation~\eqref{eq:HFInt} will be the form of the Huygens--Fresnel
integral used in the rest of the Appendix section.

As mentioned in the main text, we consider the simplest case of the
self-imaging cavity \citep{HBE-OE-08} consisting of one lens of focal
length $f$ and one mirror or radius of curvature $R$, separated
by distances $d_{1}$ and $d_{2}$, see Fig.~\ref{fig:mixsel}. The
resulting round-trip propagation matrix reads 
\begin{equation}
W=D_{1}L_{0}D_{2}MD_{2}L_{0}D_{1}\label{eq:RTO}
\end{equation}
with 
\begin{eqnarray}
D_{j} & = & \left(\begin{array}{cc}
1 & d_{j}\\
0 & 1
\end{array}\right)\,,\\
L_{0} & = & \left(\begin{array}{cc}
1 & 0\\
-\frac{1}{f_{0}} & 1
\end{array}\right)\,,\\
M & = & \left(\begin{array}{cc}
1 & 0\\
-\frac{2}{R} & 1
\end{array}\right)\,.
\end{eqnarray}
The SIC for which one finds $W=\mathrm{Id}$ is encountered for the
following values of of $d_{1,2}$ 
\begin{eqnarray}
d_{1}^{*} & = & f+\frac{f^{2}}{R}\,,\\
d_{2}^{*} & = & f+R\,.
\end{eqnarray}
 Expanding to first order in $\delta d_{j}=d_{j}-d_{j}^{\text{*}}$
one obtains 
\begin{eqnarray}
T & = & \left(\begin{array}{cc}
1 & 2\left(\delta d_{1}+\delta d_{2}\dfrac{f^{2}}{R^{2}}\right)\\
-\dfrac{2}{f^{2}}\delta d_{2} & 1
\end{array}\right)\,.
\end{eqnarray}
We conclude that, close to the SIC, small displacements from the lens
induce a $B$ coefficient, whereas moving the mirror modifies both
the values of $B$ and $C$. Finally, we can safely assume that $A=D=1$
at first order in $\mathcal{O}\left(\delta d_{j}\right)$. In addition
to small deviations from the SIC, we consider that the focal length
of the lens depends on its radial position and we model the spherical
aberration as $f\left(x\right)=f_{0}+\sigma x^{2}$. Denoting the
field profile before and after the lens as $E_{i}$ and $E_{o}$,
respectively, we have 
\begin{eqnarray}
E_{o}\left(x\right) & = & e^{-i\frac{\pi}{\lambda}\frac{x^{2}}{f}}E_{i}\left(x\right).\label{eq:LensOp}
\end{eqnarray}
Denoting the operator corresponding to the effect of the lens by $L$,
we separate the contribution from the unperturbed lens with focal
length $f_{0}$ and matrix $L_{0}$ as 
\begin{equation}
L=L_{0}+\delta L\,.\label{eq:deltaL}
\end{equation}

Using that
\begin{equation}
\frac{1}{f\left(x\right)}-\frac{1}{f_{0}}\simeq-\left(\frac{\sigma}{f_{0}^{2}}\right)x^{2}
\end{equation}
we write Eq.~\ref{eq:LensOp} as
\begin{equation}
E_{o}\left(x\right)=e^{-i\frac{\pi}{\lambda}\frac{x^{2}}{f_{0}}}\left[1+\left(e^{\frac{i\pi\sigma}{\lambda f_{0}^{2}}x^{4}}-1\right)\right]E_{i}\left(x\right),
\end{equation}
which allows to express the action of the aberration operator $\delta L:E_{i}\rightarrow E_{o}$
as 
\begin{equation}
E_{o}\simeq\left[\exp\left(i\frac{\pi}{\lambda}\frac{\sigma}{f_{0}^{2}}x^{4}\right)-1\right]\exp\left(-i\frac{\pi}{\lambda}\frac{x^{2}}{f_{0}}\right)E_{i}\,.
\end{equation}

Finally, we will employ the proximity of the SIC to simplify the representation
of the round-trip operator in the presence of the aberrations. The
full operator $W$ (cf. Eq.~\eqref{eq:RTO}) is given by 
\[
W=D_{1}LD_{2}MMD_{2}LD_{1}\,.
\]
Using Eq.~\eqref{eq:deltaL} we can expand $W$ as 
\begin{eqnarray}
W & = & W_{0}+\delta W\,,
\end{eqnarray}
where we defined the unperturbed round-trip operator as $W_{0}=D_{1}L_{0}D_{2}MD_{2}L_{0}D_{1}$.
The expression of $\delta W$ reads after simplification and to the
first order in $\mathcal{O}\left(\delta L\right)$ 
\begin{equation}
\delta W=D_{1}\left(\delta L\right)L_{0}^{-1}D_{1}^{-1}W_{0}+W_{0}D_{1}^{-1}L_{0}^{-1}\left(\delta L\right)D_{1}\,.\label{eq:dW}
\end{equation}
Further, we can simplify the dependence on $\delta L$ in Eq.~\eqref{eq:dW}
using the exact self-imaging condition which amounts to setting $W_{0}=\mathrm{Id}$
in Eq.~\eqref{eq:dW}. Indeed, since $\delta L$ is already a small
quantity, the error incurred in setting $W_{0}=\mathrm{Id}$ will
be second order. We are left calculating the two contributions of
aberrations to the round-trip field evolution. We have $W=F_{1}+F_{2}$
with 
\begin{eqnarray}
F_{1} & = & D_{1}\left(\delta L\right)L_{0}^{-1}D_{1}^{-1}\,,\,F_{2}=D_{1}^{-1}L_{0}^{-1}\left(\delta L\right)D_{1}.
\end{eqnarray}
In both cases, we can express $F_{j}$ as a double integral involving
the HFT that involve the spherical aberration of the lens. We combine
the two steps corresponding to $\gamma_{1\rightarrow2}=L_{0}^{-1}D_{1}^{-1}$
and $\gamma_{2\rightarrow1}=D_{1}^{-1}L_{0}^{-1}$ into a single Fresnel
transform. Further, we employ the SIC to find 
\begin{eqnarray}
\gamma_{1\rightarrow2} & = & \left(\begin{array}{cc}
1 & -f\left(1+\frac{f}{R}\right)\\
\frac{1}{f} & -\frac{f}{R}
\end{array}\right)\,,\\
\gamma_{2\rightarrow1} & = & \left(\begin{array}{cc}
-\frac{f}{R} & -f\left(\frac{f}{R}+1\right)\\
\frac{1}{f} & 1
\end{array}\right)\,.
\end{eqnarray}
As a last approximation, we consider the limit of a long cavity for
which the focal length of the mirror is large in comparison with the
focal length of the collimator lens. As such, we can define $\varepsilon=f/R\ll1$.
This allows to approximate the operators $\gamma_{j}$ as 
\begin{eqnarray}
\gamma_{1\rightarrow2} & = & \left(\begin{array}{cc}
1 & -f_{0}\\
\frac{1}{f_{0}} & 0
\end{array}\right)+\mathcal{O}\left(\varepsilon\right)\,,\\
\gamma_{2\rightarrow1} & = & \left(\begin{array}{cc}
0 & -f_{0}\\
\frac{1}{f_{0}} & 1
\end{array}\right)+\mathcal{O}\left(\varepsilon\right)\,.
\end{eqnarray}
Writing the action of $F_{1}$ using three integral transforms $F_{1}:E_{0}\rightarrow E_{3}$
leads to 
\begin{eqnarray}
E_{1}\left(x\right) & = & e^{i\frac{\pi}{\lambda f_{0}}x^{2}}\int_{-\infty}^{\infty}E_{0}\left(\xi\right)\sqrt{\frac{i}{\lambda f_{0}}}\exp\left[-i\frac{\pi}{\lambda f_{0}}\left(\xi-x\right)^{2}\right]d\xi,\nonumber \\
E_{2}\left(x\right) & = & \left(e^{i\frac{\pi}{\lambda}\frac{\sigma x^{4}}{f_{0}^{2}}}-1\right)e^{-i\frac{\pi}{\lambda f_{0}}x^{2}}E_{1}\left(x\right),\\
E_{3}\left(x\right) & = & \sqrt{\frac{-i}{\lambda f_{0}}}\int_{-\infty}^{\infty}E_{2}\left(\xi\right)\exp\left[i\frac{\pi}{\lambda f_{0}}\left(\xi-x\right)^{2}\right]d\xi\,.\nonumber 
\end{eqnarray}
The action of $F_{1}$ can be expressed by the following kernel 
\begin{eqnarray}
E_{3} & = & \int_{-\infty}^{\infty}E_{0}\left(z\right)K_{1}\left(x,z\right)dz\,,
\end{eqnarray}
where the kernel $K_{1}$ is defined as 
\begin{align}
K_{1}\left(x,z\right)= & \frac{1}{\lambda f_{0}}e^{-i\frac{\pi}{\lambda f_{0}}\left(z-x\right)\left(z+x\right)}\\
 & \times\int_{-\infty}^{\infty}\left(e^{i\frac{\pi}{\lambda}\frac{\sigma y^{4}}{f_{0}^{2}}}-1\right)e^{i\frac{2\pi}{\lambda f_{0}}\left(z-x\right)y}dy\,.\nonumber 
\end{align}
For small aberration, we can expand $K_{1}$ at first order in $\sigma$
which leads to 
\[
K_{1}\left(x,z\right)\simeq\frac{i\pi\sigma}{\lambda^{2}f_{0}^{3}}e^{-i\frac{\pi}{\lambda f_{0}}\left(z-x\right)\left(z+x\right)}\int_{-\infty}^{\infty}y^{4}e^{i\frac{2\pi}{\lambda f_{0}}\left(z-x\right)y}dy\,.
\]
Using $t=2\pi y/\left(\lambda f_{0}\right)$ we find 
\begin{eqnarray}
K_{1}\left(x,z\right) & = & i\frac{\sigma f_{0}^{2}}{2k^{3}}e^{-i\frac{\pi}{\lambda f_{0}}\left(z-x\right)\left(z+x\right)}\delta^{\left(4\right)}\left(z-x\right)\,,
\end{eqnarray}
where we used the fact that 
\begin{equation}
\int_{-\infty}^{\infty}t^{n}e^{i\omega t}dt=2\pi\left(-i\right)^{n}\delta^{\left(n\right)}\left(\omega\right)\,.\label{eq:deltadistr}
\end{equation}
Here, we defined $\delta^{\left(n\right)}$ as the $n^{\mathrm{th}}$
derivative of the Dirac delta which associates with the local value
of the $n^{\mathrm{th}}$ derivative of a function. Since $\delta^{\left(n\right)}$
is a well-localized function, we can set $z=x$ and obtain 
\begin{equation}
K_{1}\left(x,z\right)=i\frac{\sigma f_{0}^{2}}{2k^{3}}\delta^{\left(4\right)}\left(z-x\right)\,.
\end{equation}
Performing the convolution with $\delta^{\left(n\right)}$ is identical
to taking the fourth derivative in direct space which yields the fourth
derivative contribution. The action of $F_{1}$ to the first order
is 
\begin{eqnarray}
F_{1} & : & E_{0}\rightarrow E_{3}\,,\\
E_{3}\left(x\right) & = & i\frac{\sigma f_{0}^{2}}{2k^{3}}\partial_{x}^{4}E_{0}\left(x\right)\,.\nonumber 
\end{eqnarray}
The action of $F_{2}$ can be found in a similar fashion using three
integral transforms $F_{2}:E_{0}\rightarrow E_{3}$ and reads 
\begin{eqnarray}
E_{1}\left(x\right) & = & \int_{-\infty}^{\infty}E_{0}\left(\xi,0\right)\sqrt{\frac{-i}{\lambda f_{0}}}\exp\left[i\frac{\pi}{\lambda f_{0}}\left(\xi-x\right)^{2}\right]d\xi,\nonumber \\
E_{2}\left(x\right) & = & \left(e^{i\frac{\pi}{\lambda}\frac{\sigma x^{4}}{f_{0}^{2}}}-1\right)e^{-i\frac{\pi}{\lambda f_{0}}x^{2}}E_{1}\left(x\right),\\
E_{3}\left(x\right) & = & \sqrt{\frac{i}{\lambda f_{0}}}\int_{-\infty}^{\infty}E_{2}\left(\xi\right)\exp\left[-i\frac{\pi}{\lambda f_{0}}\left(-2x\xi+x^{2}\right)\right]d\xi\,.\nonumber 
\end{eqnarray}
Similarly, the action of $F_{2}$ can be expressed as a integral
\begin{equation}
E_{3}=\int_{-\infty}^{\infty}E_{0}\left(z,0\right)K_{2}\left(x,z\right)dz\,,
\end{equation}
where we defined the kernel $K_{2}$ as 
\begin{eqnarray}
K_{2}\left(x,z\right) & = & \frac{1}{\lambda f_{0}}e^{i\frac{\pi}{\lambda f_{0}}\left(z-x\right)\left(z+x\right)}\\
 & \times & \int_{-\infty}^{\infty}\left(e^{i\frac{\pi}{\lambda}\frac{\sigma y^{4}}{f_{0}^{2}}}-1\right)\exp\left[i\frac{2\pi}{\lambda f_{0}}y\left(x-z\right)\right]dy\,.\nonumber 
\end{eqnarray}
Expanding analogously the aberration contribution to first order in
$\sigma$ and using Eq.~\eqref{eq:deltadistr}, we obtain the desired
kernel 
\begin{eqnarray}
K_{2}\left(x,z\right) & = & i\frac{\sigma f_{0}^{2}}{2k^{3}}\delta^{\left(4\right)}\left(z-x\right)\,.
\end{eqnarray}
We conclude that, to the first order in $\sigma$, the action of $F_{2}$
is identical to that of $F_{1}$
\begin{eqnarray}
F_{2}:E_{0} & \rightarrow & E_{3}\,,\\
E_{3}\left(x\right) & = & i\frac{\sigma f_{0}^{2}}{2k^{3}}\partial_{x}^{4}E_{0}\left(x\right)\,.\nonumber 
\end{eqnarray}

In summary, the total effect due to spherical aberration close to
self-imaging in the one-dimensional case reads 
\begin{equation}
E_{3}\left(x\right)=i\frac{\sigma f_{0}^{2}}{k^{3}}\partial_{x}^{4}E_{0}\left(x\right)\,.
\end{equation}
The generalization of these calculations to two-dimensions proceeds
without difficulties and yields
\begin{equation}
E_{3}\left(r_{\perp}\right)=i\frac{\sigma f_{0}^{2}}{k^{3}}\nabla_{\perp}^{4}E_{0}(r_{\perp})\,.
\end{equation}


\end{document}